\documentclass[preprint,superscriptaddress,showpacs,preprintnumbers,amsmath,amssymb,aps,pre]{revtex4-2}
\usepackage[utf8]{inputenc} 
\usepackage{epsfig}
\usepackage{natbib}
\usepackage[T1]{fontenc}

\usepackage{booktabs}
\usepackage{amsmath}
\usepackage{times}
\usepackage{epstopdf} 
\usepackage{graphicx}
\usepackage{color}

\usepackage{afterpage}

\begin{document}

\title{Trapping enhanced by noise in nonhyperbolic and hyperbolic chaotic
scattering}

\author{Alexandre R. Nieto}
\email[]{alexandre.rodriguez@urjc.es}
\affiliation{Nonlinear Dynamics, Chaos and Complex Systems Group, Departamento de
F\'{i}sica, Universidad Rey Juan Carlos, Tulipán s/n, 28933 M\'{o}stoles, Madrid, Spain}

\author{Jes\'{u}s M. Seoane}
\affiliation{Nonlinear Dynamics, Chaos and Complex Systems Group, Departamento de
F\'{i}sica, Universidad Rey Juan Carlos, Tulip\'{a}n s/n, 28933 M\'{o}stoles, Madrid, Spain}

\author{Miguel A.F. Sanju\'{a}n}
\affiliation{Nonlinear Dynamics, Chaos and Complex Systems Group, Departamento de
F\'{i}sica, Universidad Rey Juan Carlos, Tulip\'{a}n s/n, 28933 M\'{o}stoles, Madrid, Spain}
\affiliation{Department of Applied Informatics, Kaunas University of Technology, Studentu 50-415, Kaunas LT-51368, Lithuania}

\date{\today}

\begin{abstract}

The noise-enhanced trapping is a surprising phenomenon that has already been studied in chaotic scattering problems where the noise affects the physical variables but not the parameters of the system. Following this research, in this work we provide strong numerical evidence to show that an additional mechanism that enhances the trapping arises when the noise influences the energy of the system. For this purpose, we have included a source of Gaussian white noise in the Hénon-Heiles system, which is a paradigmatic example of open Hamiltonian system. For a particular value of the noise intensity, some trajectories decrease their energy due to the stochastic fluctuations. This drop in energy allows the particles to spend very long transients in the scattering region, increasing their average escape times. This result, together with the previously studied mechanisms, points out the generality of the noise-enhanced trapping in chaotic scattering problems.

\end{abstract}

\maketitle
\newpage
\section{Introduction} \label{sec:Introduction}
Chaotic scattering in open Hamiltonian systems is an important topic
in nonlinear science due to its fundamental applications in
classical \cite{Seoane13,Lai,Telb} and quantum physics
\cite{Lu,Ott,Stock}. This phenomenon is also relevant in a wide
variety of fields such as chemistry \cite{Ezra,Kawai}, biology
\cite{Tel,Scheuring} and even medicine \cite{Schelin}. Most of the
work in chaotic scattering has been made by using purely
conservative systems
\cite{Bleher,Contopoulos93,Kandrup99,Aguirre01}. However, almost
every system in nature is influenced by its surrounding environment.
For that reason, recently some research has considered perturbations
in the conservative systems as an attempt to model the coupling of
the system with the environment. In particular, the effects of
dissipation \cite{MotterLai,MotterLai2,SeoaneLai2}, periodic forcing
\cite{Blesa14,Nieto18} and noise \cite{Silva,Rodrigues,Altmann,
Bernal} have been considered. It has been shown that weak
dissipation can convert the Kolmogorov-Arnold-Moser (KAM) islands
into attractors, and that the periodic forcing can destroy the KAM
structures at the same time that it decreases the unpredictability of
the system \cite{Nieto18}.

Regarding the effects of noise, much work has been done in the
recent years. Most of this research has focused on chaotic maps where an additive noise affects the physical variables of the systems. In this situation it has been shown that the noise can destroy the small scales of the KAM islands
\cite{Mills}, leading to escapes that would be forbidden in the
deterministic system \cite{Silva,Rodrigues}. Nevertheless, the
opposite effect can occur and trajectories that do not belong to the
KAM islands can enter inside them and describe a
transient regular motion \cite{Rodrigues,Altmann}. Furthermore, another interesting phenomenon has been reported: the noise-enhanced trapping \cite{Altmann}. Small noise
intensities can play a constructive role by reducing the escape rate
of the particles. The noise-enhanced trapping  appears in both fully
chaotic (hyperbolic) and mixed-phase-space (nonhyperbolic) systems,
even if the mechanism that generates it is different in each case. In
the fully chaotic case, the escape rate is reduced due to the blurring of
the natural measure of the exits, and therefore allowing the
trajectories to avoid their escape. In mixed-phase-space systems the
trapping is enhanced by means of some trajectories that, even
starting inside the chaotic sea, can enter the regions defined by
the KAM islands in the noiseless case.

Regarding the effect of noise in continuous open Hamiltonian
systems, we find less examples in the literature. In particular, one
relevant work (see Ref.~\cite{Bernal}) in which a threshold value of
the noise intensity changes the decay law from algebraic to
exponential in mixed-phase-space systems has been uncovered. The
exponential decay remains ubiquitous for strong noise intensities
\cite{Seoane08}. In addition, some investigations of the effects of
the noise into the fractal structures of the exit basins has been
carried out, both in presence of Gaussian noise \cite{Bernal}
and noisy periodic forcing \cite{Gang}. However, as far as we know,
the noise-enhanced trapping has not been the focus of attention in
the context of the continuous Hamiltonian systems.

The main goal of our manuscript is to show that an additional mechanism to those described above takes place in open Hamiltonian systems. At the same time, we also confirm that the previously studied mechanism which occurs in mixed-phase-space maps is also relevant when the energy of the system is subject to fluctuations. Since most of chaotic maps are obtained as a Poincaré surface of section of continuous systems, the results are expected to be identical in both kind of systems. However, in open Hamiltonian systems the introduction of noise in the momentum or in the spatial coordinates indirectly affects the energy. This main difference allows a new and supplementary mechanism to occur. Surely, this mechanism is expected to play a certain role in chaotic maps where the noise affects the parameters of the system (e.g. multiplicative noise).

The mechanism that we have undercover appears in both hyperbolic and non-hyperbolic regimes and it is related to the possibility that stochastic fluctuations allow some trajectories, that we call \textit{unusual}, to remain in the scattering region with low energy values during very long transients. The smaller the energy, the smaller the exit set of the system and, in consequence, higher
the escape times.Moreover, for low energy values the phase space of many systems is occupied by extensive KAM tori, so the noise that enhances the trapping allows the trajectories to enter in KAM regions of the deterministic system that exist for different values of the energy. In this sense, the new mechanism reported in this manuscript can be understood as a catalyst of the previously investigated one.
	
	To study this issue, we have chosen the Hénon-Heiles
system \cite{HH64}, which represents a paradigmatic example of
chaotic scattering that exhibits a transition between hyperbolic and
nonhyperbolic dynamics. This characteristic is interesting because
we can analyze the noise-enhanced trapping in fully chaotic and
mixed-phase-space cases using the same model.

The structure of this paper is as follows. In Sec.~\ref{Model}, we
describe our model, the H\'{e}non-Heiles system with additive
uncorrelated Gaussian noise. In Sec.~\ref{Chaotic} and
Sec.~\ref{Mixed} we show strong numerical evidence of noise-enhance
trapping in the fully chaotic and mixed-phase-space regimes of the
system, respectively. In the latter, we also show the gradual
reduction of the stickiness of the KAM islands under the effect of very weak noise.
The explanation about the mechanism leading to the trapping is
carried out in Sec.~\ref{Mechanism}. Finally, in
Sec.~\ref{Conclusions}, we present the main conclusions of this
manuscript.

\section{Model description} \label{Model}

\begin{figure}[htp]
    \centering
    \includegraphics[width=0.45 \textwidth,clip]{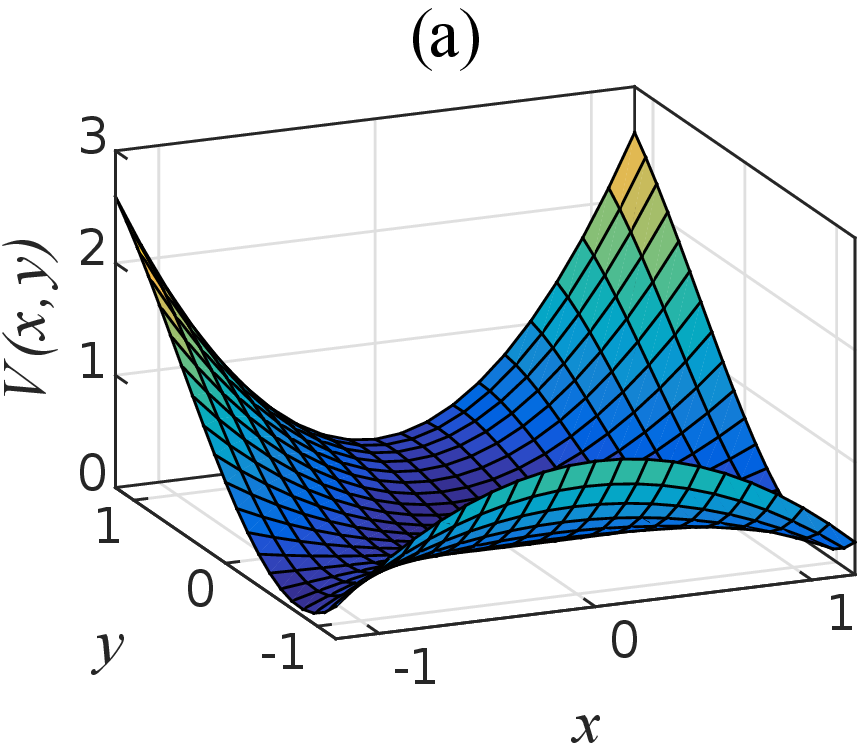} \hspace{.5 cm}
    \includegraphics[width=0.45\textwidth,clip]{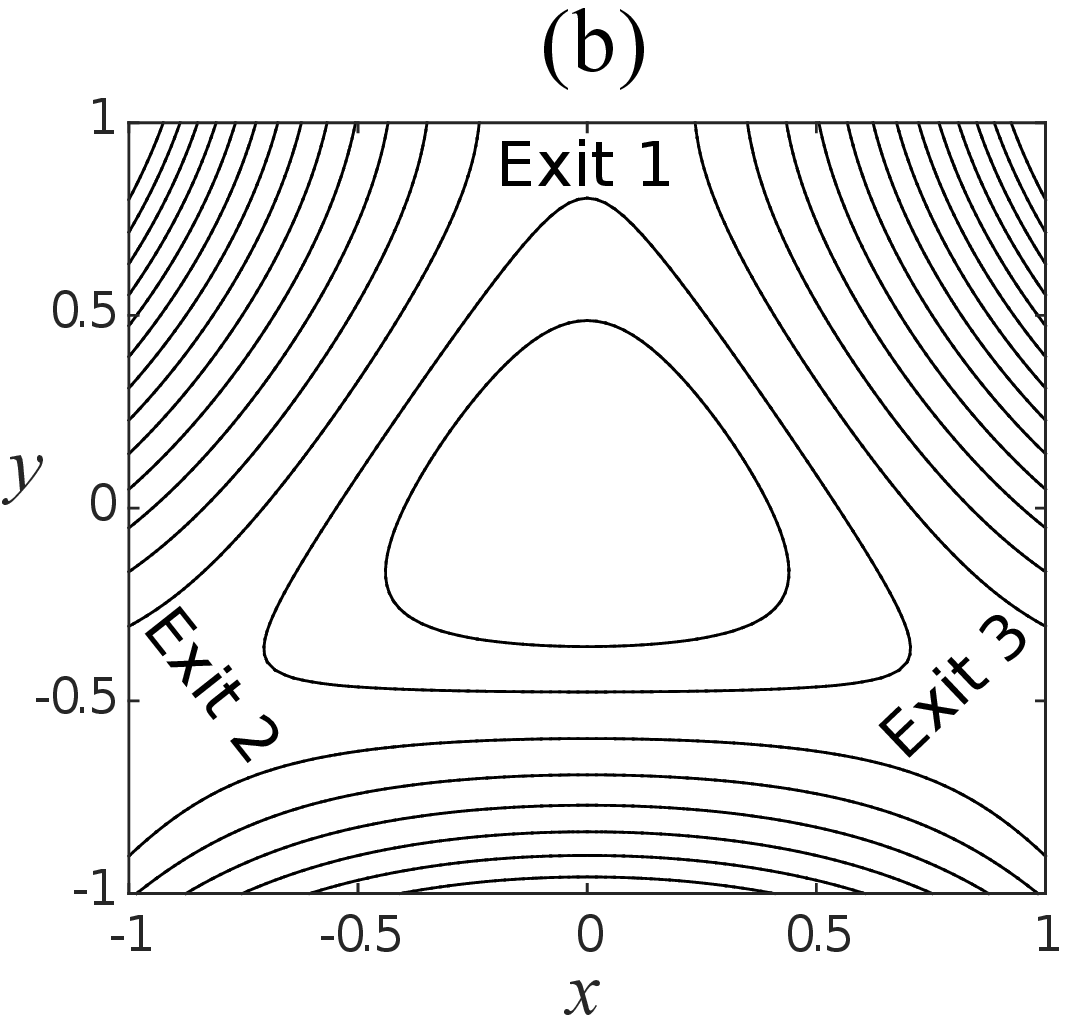}
    \caption{(a) Potential $V(x,y)$ of the the H\'{e}non-Heiles system. (b) Isopotential curves for different values of the energy, under and over the threshold value $E_e=1/6$. For energy values $E>E_e$ three different exits are visible: exit 1 ($y\to\infty$), exit 2 ($x,y\to-\infty$), and exit 3 ($x\to\infty,y\to-\infty$).}
    \label{fig:pot}
 \end{figure}

The model that we have used to study the noise-enhanced trapping in
continuous open Hamiltonian systems is the Hénon-Heiles.
This Hamiltonian system appeared in the literature for the first
time in $1964$ as a model of a particle moving in a galactic
potential. The Hamiltonian is given by the sum of the kinetic energy
and a nonlinear axisymmetric potential [see Fig.~\ref{fig:pot}(a)]:

\begin{equation} \label{eq:HH_Hamiltonian}
{\cal{H}}=\frac{1}{2}(p^2+q^2)+\frac{1}{2}(x^2+y^2)+x^2y-\frac{1}{3}y^3,
\end{equation}
where $x$ and $y$ are the coordinates, and $\dot{x}=p$ and $\dot{y}=q$ denote the two components of the generalized momentum.

The system has been extensively studied due to the wealth of
dynamical behaviors that exhibits for different values of the energy
\cite{Barrio08,Barrio09,Vallejo03,ZotosHH}. Particularly, the
system exhibits two energy thresholds that separate very different
regimes. The first threshold occurs when the energy is $E_e=1/6$.

For energy values below $E_e$ the isopotential curves are close,
while for energy values over $E_e$ three exits separated by an angle
of $2\pi/3$ appear. Then the particles can escape from the
scattering region to infinity and the system becomes an open
Hamiltonian system. To visualize this fact, we depict in
Fig.~\ref{fig:pot}(b) the isopotential curves for different values
of the energy, under and over $E_e$. For energy values slightly
higher than $E_e$ the mixed-phase-space is occupied by enormous KAM
islands that make the dynamical behavior of the system
nonhyperbolic. The nonhyperbolic character manifests, among other
ways, through an algebraic decay law in the survival probability of
the particles. If we increase the energy more and more, the size of
the KAM islands decreases irregularly until $E_t \simeq 0.23$
\cite{Nieto20}, which is the second important threshold. Over $E_t$
all the KAM tori have been destroyed and the hyperbolic regime
starts. This regime is characterized by an exponential decay law and
a fully chaotic phase space.

For the purposes of this research we include in the Hénon-Heiles
system a source of additive uncorrelated Gaussian noise. Under this
consideration the equations of motion read \cite{Seoane08}:

\begin{equation} \label{eq:eq motion}
\begin{aligned}
\dot{p} & = -x - 2xy + \sqrt{2\xi}\eta_x(t) \\
\dot{q} &= -y - x^2 + y^2 + \sqrt{2\xi}\eta_y(t),
\end{aligned}
\end{equation}
where $\xi$ is the intensity of the noise and $\eta_x(t)$,$\eta_y(t)$ are Gaussian white noise processes with mean $\mu=0$ and variance
$\sigma^2 = 2\xi$.

To solve numerically this system of stochastic differential
equations we have used the stochastic second-order Heun method
\cite{Kloeden}, as was previously used in
Refs.~\cite{Bernal,Seoane08}. To ensure the effectiveness of the
method we have tested the stability of the solutions and the
convergence of the main magnitudes of this research (average escape
time and energies). Since the Heun method is not a symplectic
algorithm and we have worked with a conservative system, we have
included a negligible amount of dissipation in order to avoid
stability problems in the integration algorithm. We have also
compared the results by using the numerical schemes of the
Euler-Maruyama method and the stochastic fourth-order Runge-Kutta
method \cite{Kloeden}. All the three methods have provided similar
results with a good precision. However, the Heun method has been the
one that offers the best balance between precision and
computational effort.

\section{Noise-enhanced trapping in the fully chaotic regime} \label{Chaotic}

In this section, we study the phenomenon of noise-enhanced trapping
in the fully chaotic Hénon-Heiles system. As we commented in the
introduction, the last significant KAM torus is destroyed at the energy level
$E_t=0.23$. However, it has been shown that the particles that belong
to KAM islands are completely residual for energies above $E=0.21$
\cite{Nieto20} and the phase space can be considered fully chaotic.
For that reason, we have decided to consider in this section energies
$E>0.21$ instead of $E>0.23$. The results of the next section will
show that this division of energies is natural when studying the
effect of noise in this system.

The main magnitude that we have studied to characterize the
dynamical behavior is the average escape time of the trajectories,
$T$. Because the system is $4$-dimensional and we have to deal with
$3$ free coordinates, there exist many methods to launch trajectories
in phase space in order to obtain $T$. Some researchers fix the
physical coordinates as $(x,y)=(0,0)$ and then vary the angle
of the launching. Other authors prefer to fix the angle and vary the
physical coordinates over a line segment in the physical space. In
order to consider a more general range of initial conditions in phase space, here we have used the tangential shooting method
\cite{Aguirre01}. Using this method, the velocity vector is defined
in such a way that is tangent to the circle centered at the origin
and passing through the point $(x_0,y_0)$. This consideration
implies that when we vary $(x_0,y_0)$ we are also varying
$(p_0,q_0)$, so we are exploring initial conditions in all the phase
space. To choose a suitable method to define the initial conditions
is important, especially in the mixed-phase-space-regime, because it is
possible to miss some parts of the dynamical behavior (trapped
trajectories for example) by choosing specific sets of initial
conditions.

\begin{figure}[htp]
    \centering
    \includegraphics[width=0.49\textwidth,clip]{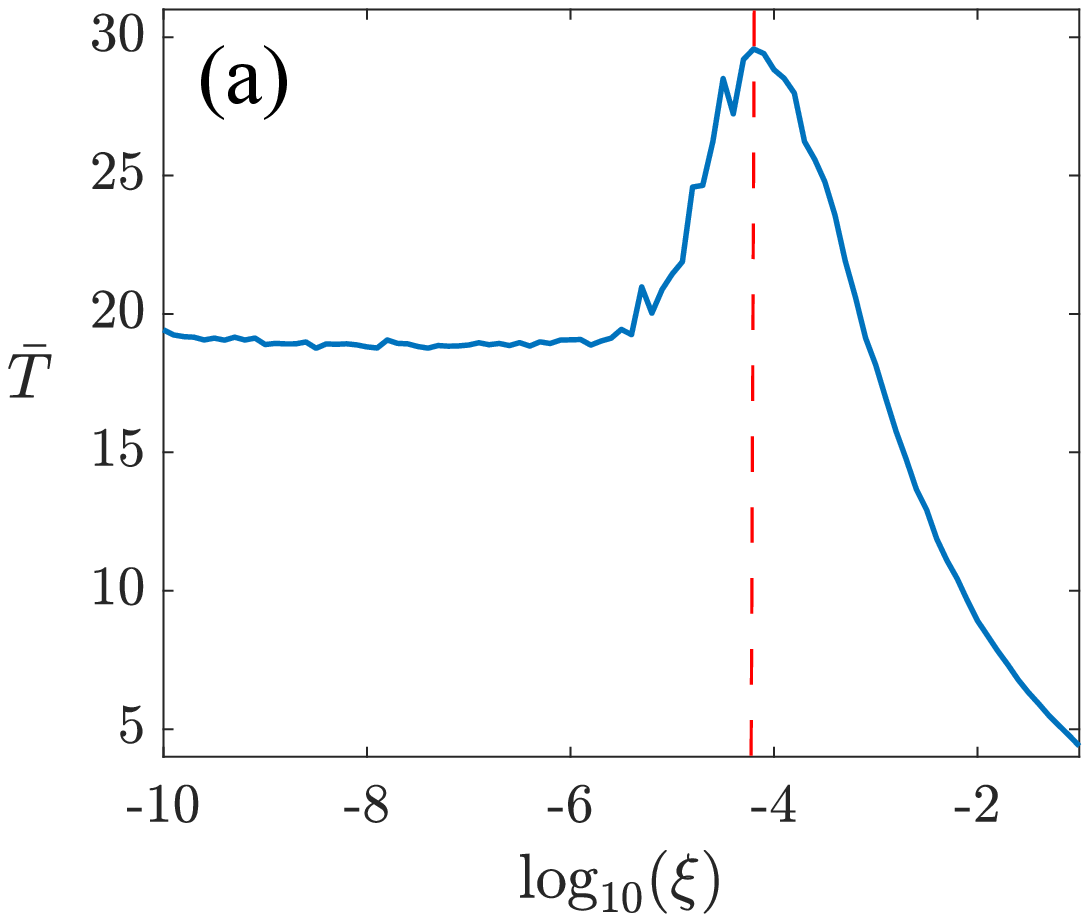}
    \includegraphics[width=0.49\textwidth,clip]{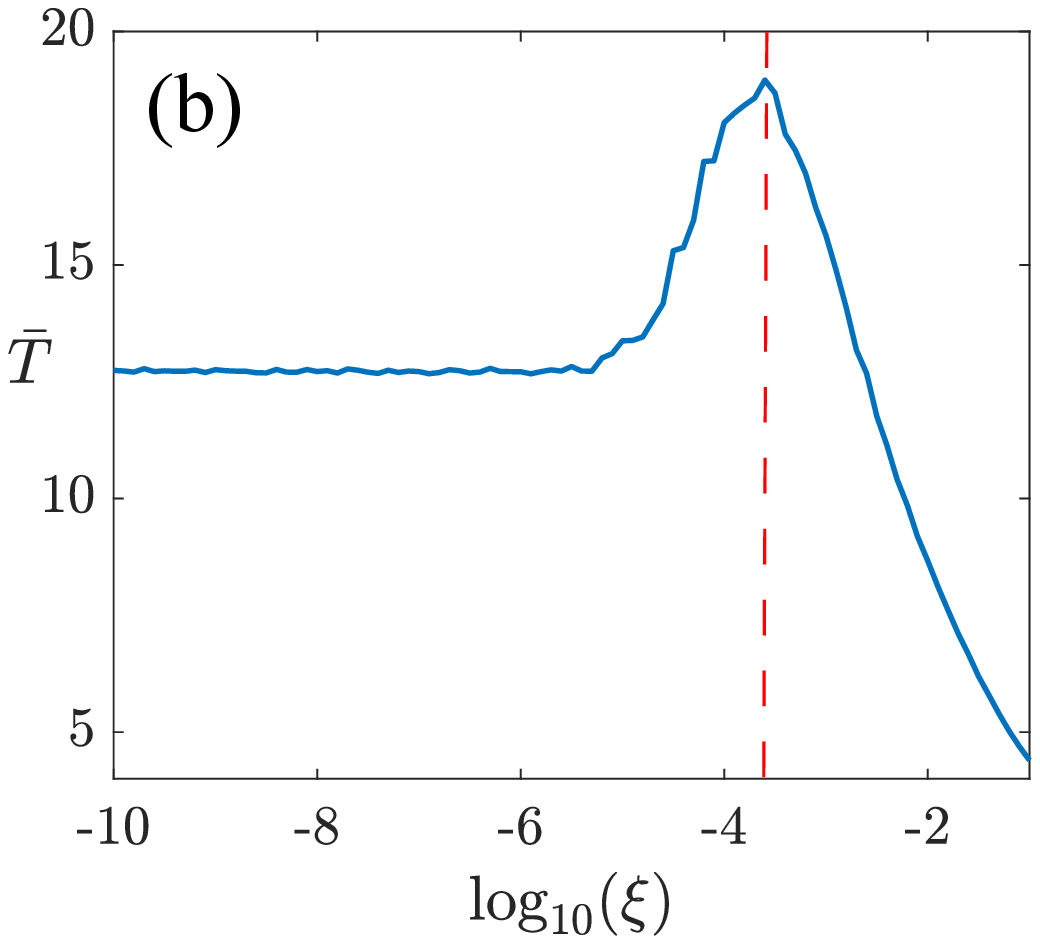}
    \caption{Evolution of the average escape time with increasing noise for (a) $E=0.23$ and (b) $E=0.25$. The red dashed lines are located at the maximum of the average escape time. In order to compute this figure, we have used $200$ values of $\xi\in[10^{-10},10^{-1}]$. For every noise intensity, we have launched $250000$ initial conditions and calculated the average escape time.}
    \label{Tmax_hy}
\end{figure}

Now, we are going to describe the relevant results of this
section. We have considered a wide range of noise intensities
$\xi\in[10^{-10},10^{-1}]$ and computed the average escape times of
the particles for energies $E=0.23$ and $E=0.25$. The results are
shown in Fig.~\ref{Tmax_hy}, where we can observe a maximum in
the average escape time, showing that for specific values of the
noise, $\xi_t$, the trapping is enhanced. For very weak noise
intensities the escape times remain unaltered, while for very strong
noises the escape time decreases abruptly.

\begin{figure}[htp]
    \centering
    \includegraphics[width=0.55\textwidth,clip]{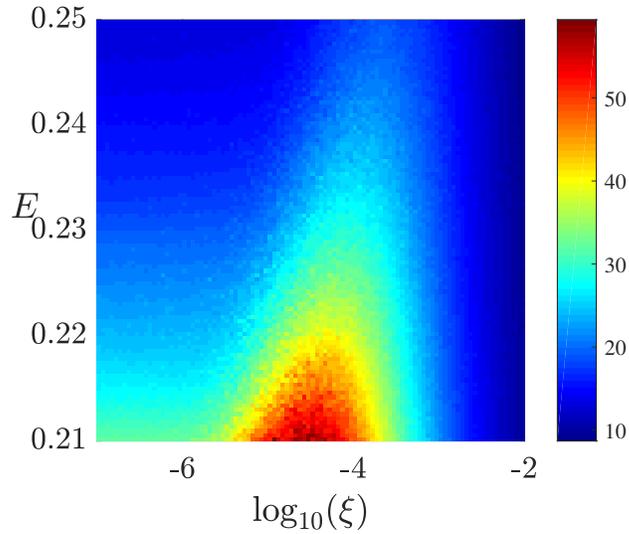}
    \caption{Color-code map showing the average escape times for several values of the energy and the noise intensity. The hot colors indicate high values of the escape times. We have used $200\times 200$ equally spaced values of the parameters. For every combination of energy and noise we have computed $50000$ trajectories in order to calculate the average escape time. The figure clearly shows a noise intensity that allows the particles to describe unusual long transients in the scattering region. }
    \label{pcolor_hy}
\end{figure}

To show these results in a more general manner, we have computed the
average escape times for $40000$ combinations of energies
$E\in[0.21,0.25]$ and noise intensities $\xi\in[10^{-7},10^{-2}]$.
To visualize the result we have used in Fig.~\ref{pcolor_hy}, a
color-coded map in which hot colors indicate high average escape
times. As we could also observe in Fig.~\ref{Tmax_hy}, this
color-coded map shows that the value of $\xi_t$ increases with the
energy, at the same time that the intensity of the trapping is reduced. This implies that if the energy is increased to values deep in the fully chaotic regime, the effect of the trapping in the average escape times might be negligible. In this situation the traditional mechanism of trapping \cite{Altmann}, based on a reduction of the escape rate due to the blurring of the natural measure of the system, is expected to dominate the trapping phenomenon of the particles in the scattering region.

In chaotic maps with leaks, the noise enhances the trapping
by blurring the measure of the leak $I$. The noise can move
trajectories out of $I$, so they do not escape in the next
iteration. In the case of continuous open Hamiltonian systems, the
exit set is defined by the highly unstable periodic orbits called Lyapunov
orbits \cite{Contopoulos90}. Once a particle cross a Lyapunov orbit
with the velocity vector pointing out, low noise intensities
cannot make the particle to climb the hill and return to the
scattering region. In particular, for energies $E\in[0.23,0.25]$ our simulations have established
that only noise intensities $\xi>0.1$ allow some trajectories to
return to the scattering region after crossing slightly a Lyapunov
orbit.

\section{Noise-enhanced trapping in the mixed-phase-space regime} \label{Mixed}

In this section we analyze the noise-enhanced trapping in the
mixed-phase-space regime of the system. We have started our
computations in $E=0.18$, even if the exits are open over $E_e=1/6$.
We have decided this to avoid the immense computational effort that
is necessary when dealing with the convergence of the trajectories
for energy values very close to $E_e$.

\begin{figure}[htp]
    \centering
    \includegraphics[width=0.8\textwidth,clip]{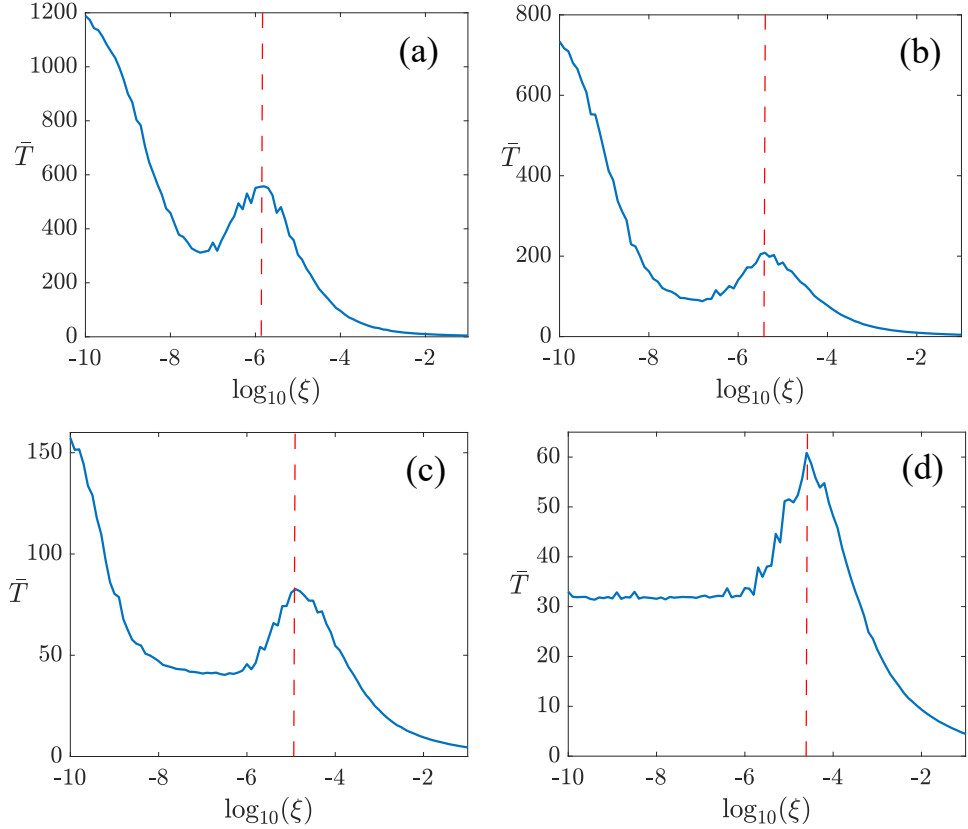}
    \caption{Evolution of the average escape time with increasing noise for (a) $E=0.18$, (b) $E=0.19$, (c) $E=0.20$ and (d) $E=0.21$. The red dashed lines are located at the relative maxima of the average escape time. In order to compute this figure we have used $200$ values of $\xi\in[10^{-10},10^{-1}]$. For every noise intensity we have launched $250000$ initial conditions and calculated the average escape time. The maximum integration time that we have considered is $t_{max}=100000$. In (a-c) panels a decrease in $T$ due to the progressive destruction of the stickiness of the KAM islands is observed. The average escape time of the deterministic system are $T_a=1314$, $T_b=792$, $T_c=161$, and $T_d=33$.}
    \label{net_mps}
\end{figure}

In Fig.~\ref{net_mps} we show the evolution of the average escape
time in function of the noise intensity for four energy values
within the nonhyperbolic regime. Here, we can observe local maxima
associated with the trapping for noise intensities between $10^{-4}$
and $10^{-6}$. The main qualitative difference between this figure
and Fig.~\ref{Tmax_hy} is that an initial decrease in $T$ appears
for energies $0.18$, $0.19$ and $0.20$. As we will show, this
decrease is directly related to the reduction of the stickiness of the KAM islands
due to he effects of noise. In the case $E=0.21$ no decrease is
observed due to the small size and influence of the KAM islands.

\begin{figure}[htp]
    \centering
        \includegraphics[width=0.4\textwidth,clip]{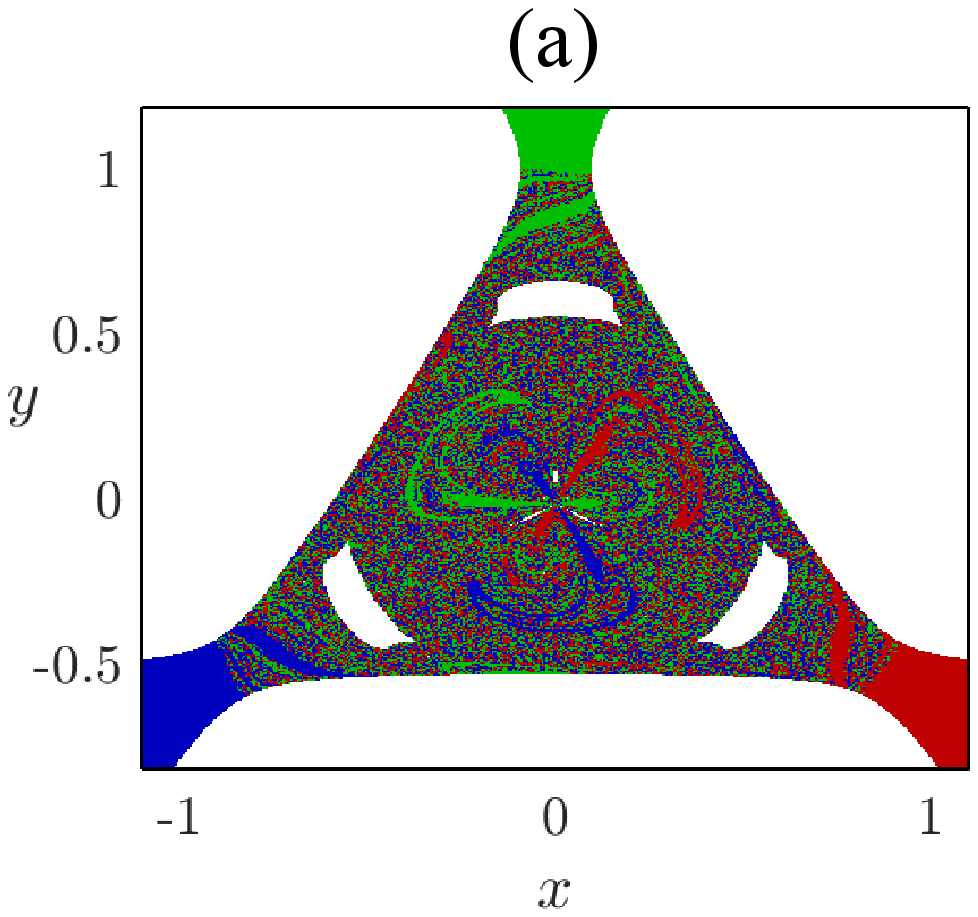}
    \includegraphics[width=0.4\textwidth,clip]{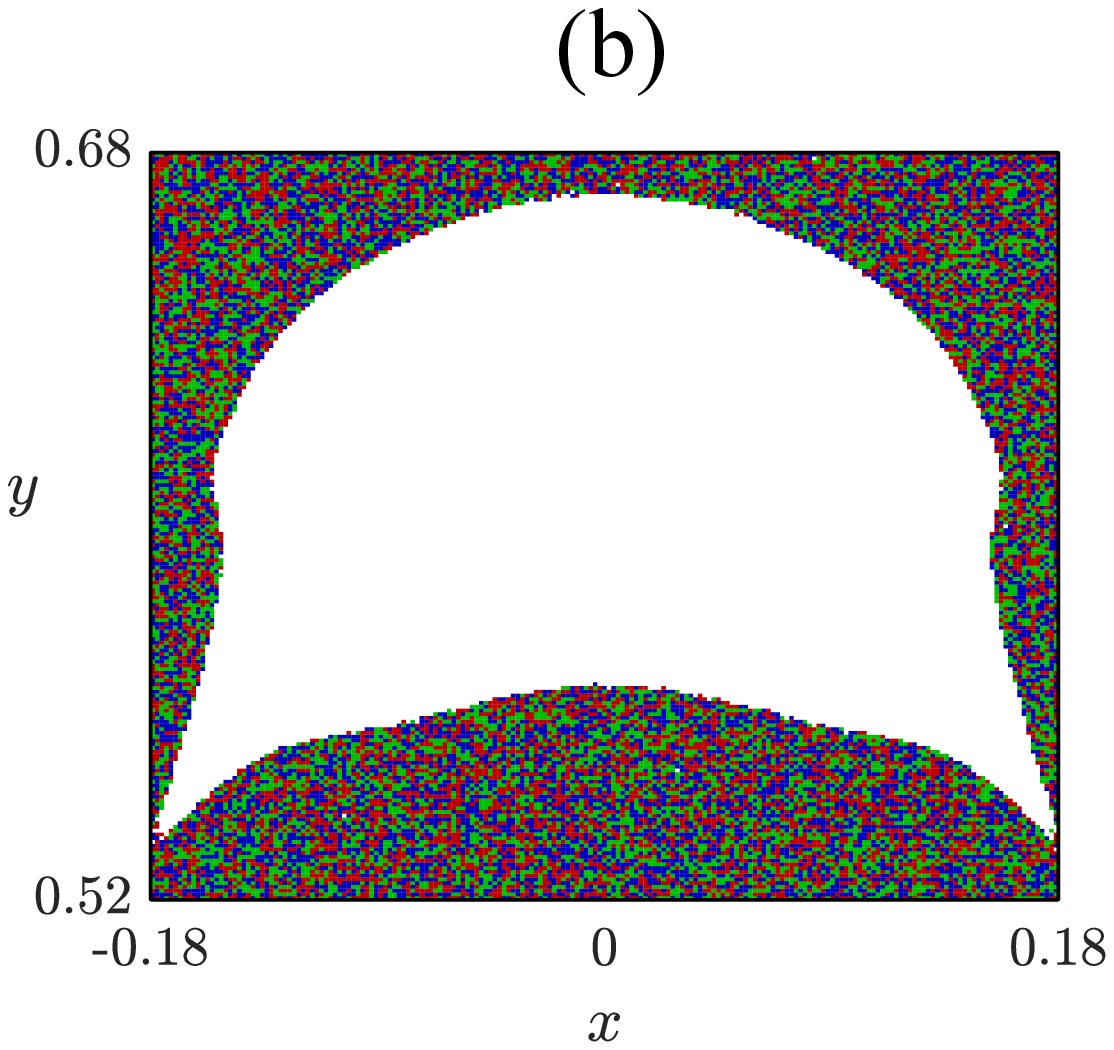}
    \caption{(a) Exit basins in the physical space $(x,y)$ for the Hénon-Heiles system with energy $E=0.18$. The colors green, blue and red correspond to initial conditions escaping through exists $1$, $2$ and $3$, respectively. The white color refers to particles that belong to a KAM torus and never escape. (b) Zoom-in on the KAM island which is close to the exit $1$. To compute this figure we have used (a) $400\times400$ and (b) $200\times200$ initial conditions. }
    \label{exitbasins}
\end{figure}

Strictly speaking, KAM islands do not exist in presence of noise. However, their ghosts manifest through their stickiness which can retain the particles during long transients. As the noise intensity is increased the stickiness is reduced, so the higher the noise the higher the probability of escaping from a KAM region in a certain time. A good way to visualize this reduction in the stickiness is through the exit basins \cite{Contopoulos02}. In the context of continuous open Hamiltonian systems, we define an exit basin as the set of initial conditions that after a finite time will escape through a certain exit. In order to visualize them in a figure, we establish a different color to the initial conditions depending on the exit through which they will escape. We use another color, usually black or white, to represent the trajectories that belong to a KAM torus and do not escape. In Fig.~\ref{exitbasins} we represent (a) the exit basins in the physical space $(x,y)$ for $E=0.18$, and (b) a zoom-in in the exit basins, centered in one of the main three KAM islands of the deterministic system.

\begin{figure}[htp]
	\centering
	\includegraphics[width=1\textwidth,clip]{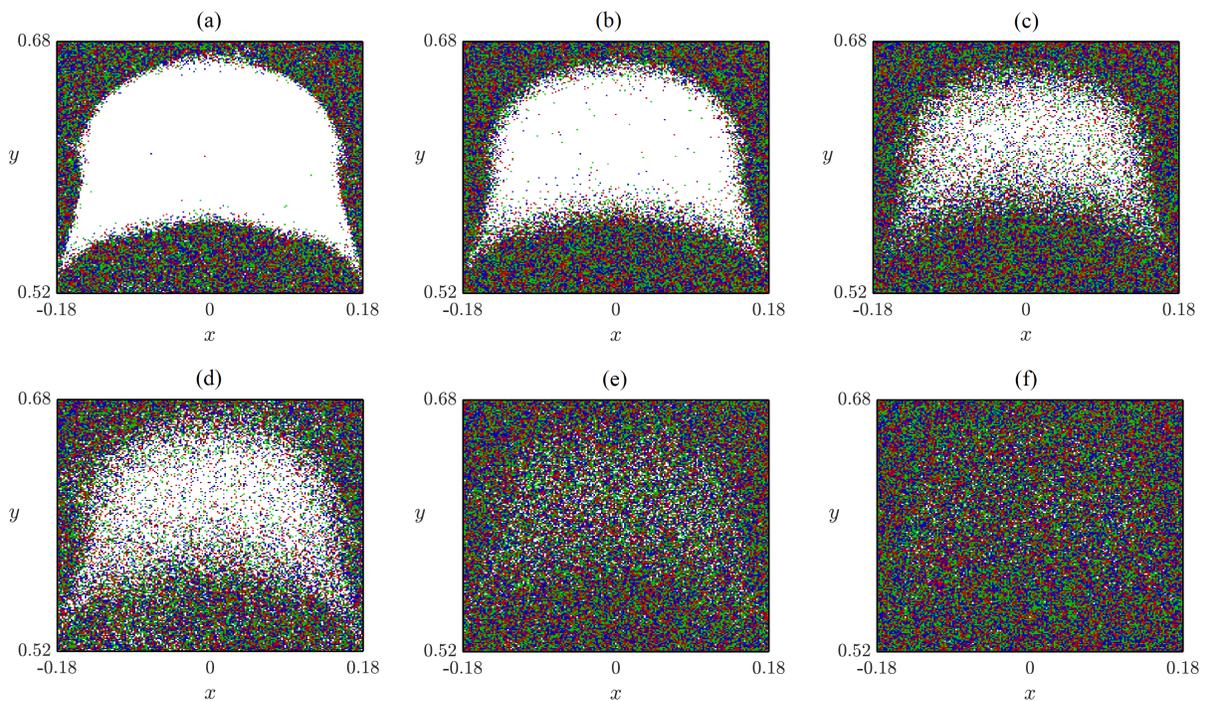}
	\caption{Zoom-in of the exit basins of the Hénon-Heiles system with energy $E=0.18$.
			The intensities of the noise are $\xi_a=\xi_b=\xi_c=10^{-9}$ and $\xi_d=\xi_e=\xi_f=5\times10^{-8}$. The maximum integration time is $t_{a}=t_{d}=1~000$, $t_{b}=t_{e}=5~000$ and $t_{c}=t_{f}=20000$. The structure of the KAM regions is progressively blurred when increasing both the noise intensity and the maximum integration time. We have checked in all the cases that there are no trapped particles after $t_{max}=50000$. }
	\label{kam_blurr}
\end{figure}

The effect of a source of noise weaker than the value that enhances the trapping is to allow particles inside
the KAM islands to leave the region and escape, so a blurring in the
KAM regions in the exit basins is expected. However, the intensity
of the blurring that we observe depends on the maximum integration
time that we consider. In Fig.~\ref{kam_blurr} we show the same
zoom-in of Fig.~\ref{exitbasins}(b), for different noise intensities
and different maximum integration times. By observing the pictures
from left to right (increasing $t_{max}$) it is clear that the KAM
islands are gradually blurred. In fact, we have tested that for
$t_{max}=50000$ there are no trapped particles in phase space.
On the other hand, by observing the pictures from top to bottom
(increasing the noise) we detect a faster blurring for higher noise.
The higher the noise, the greater the reduction of the stickiness of
the KAM islands, which generates the decreasing in the average
escape times that we have observed in Fig.~\ref{net_mps}.

\begin{figure}[htp]
    \centering
\includegraphics[width=1\textwidth,clip]{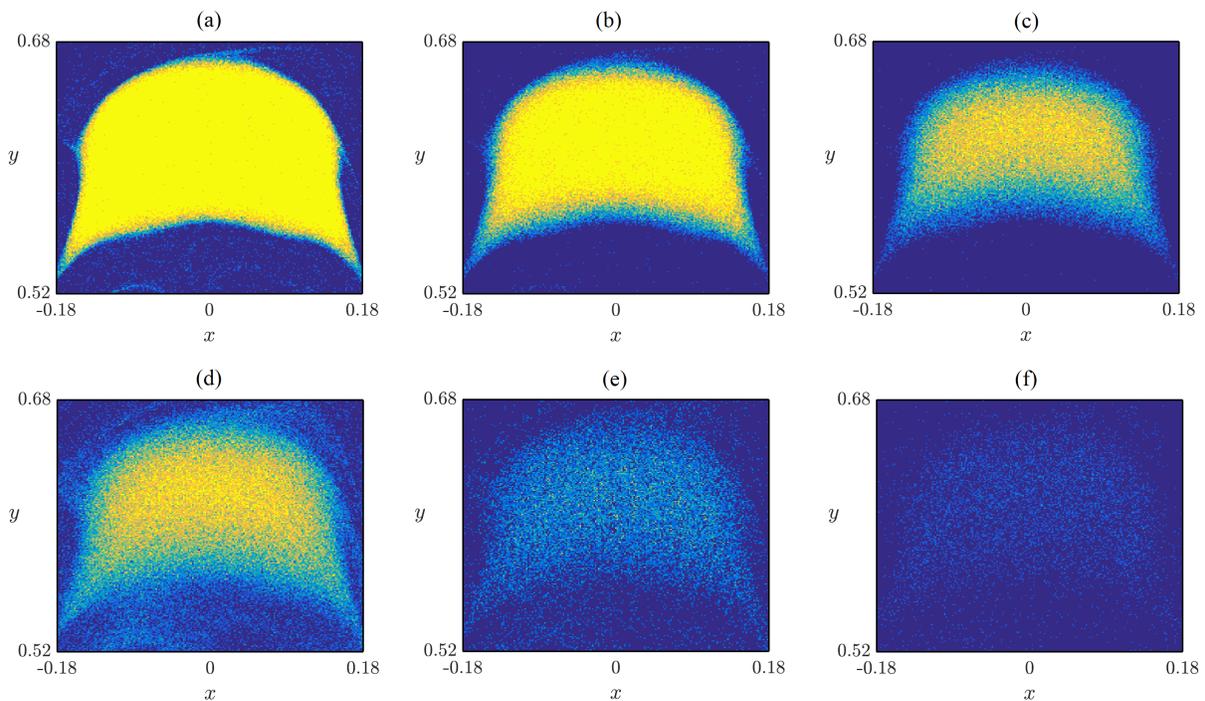}
    \caption{Color-code showing the probability that an initial condition close to the KAM regions in the physical space for $E=0.18$ remains in phase space once the maximum integration time is reached. Yellow color means probability $1$, dark blue means probability $0$, and the other colors refer to intermediate probabilities. The intensity of the noise is $\xi_a=\xi_b=\xi_c=10^{-9}$ and $\xi_d=\xi_e=\xi_f=5\times10^{-8}$. The maximum integration time is $ t_a=t_d=1~000$, $t_b=t_e=5000$ and $t_c=t_f=20000$. To generate this figure we have used $100$ exit basins and computed the probability for every initial condition. It is clear that the probability decreases with both the maximum integration time and the noise.}
    \label{fig:basinsHH_prob}
\end{figure}

Even if Fig.~\ref{kam_blurr} is clear enough for its purpose, it
is worth noting that the exit basins have no meaning when
working with noisy systems. Due to the effect of noise in the
chaotic behavior, the same initial condition can escape through a
different exit in different launchings. This implies that the color
that appear in the exit basins can be different in different
simulations. Being rigorous, the appropriate representation of the
asymptotic behavior is by using probabilities. In particular, we can
generate figures similar as Fig.~\ref{kam_blurr} many times (say
$100$) and calculate the probability that a particle remains in the
KAM island after the maximum integration time. This is exactly what
we show in Fig.~\ref{fig:basinsHH_prob}, where the yellow color means
probability $1$, while dark blue means probability $0$. Intermediate
colors imply intermediate probabilities. From this figure, we can
state similar conclusions as in Fig.~\ref{kam_blurr}. For
$\xi=10^{-9}$ the boundary of the KAM islands starts to have the
possibility to escape when we increase the maximum integration time.
In the case of $\xi=5\times10^{-8}$ almost all initial conditions
have an option to escape for $t_{max}=5000$, while for
$t_{max}=50000$ the probability to remain in the scattering region
is almost zero for every initial condition. This figure serves just as another evidence of the
gradual reduction of the stickiness for weak noise.

The fact that the stickiness of the KAM islands is reduced with increasing noise does not imply that they do not affect the escape dynamics in the noise-enhanced trapping. On the contrary, as it has been shown in previous works \cite{Altmann}, due to the effects of noise, particles that describe escaping orbits in the deterministic system can jump into the KAM islands describing long transients. Hence, the initial decrease on the average escape time is explained by the reduction of the stickiness, but the relative maximum (i.e. the noise-enhanced trapping) is strongly related to the KAM structures. In continuous open Hamiltonian systems, changes in the physical variables due to the noise imply changes in the energy, so the particles are able to enter in the structure of the existing KAM islands for different values of the energy. However, this does not happen for every noise intensity, but only for the intensities that suit the characteristics of the fine structure of the KAM islands. 

\begin{figure}[htp]
	\centering
	\includegraphics[width=1\textwidth,clip]{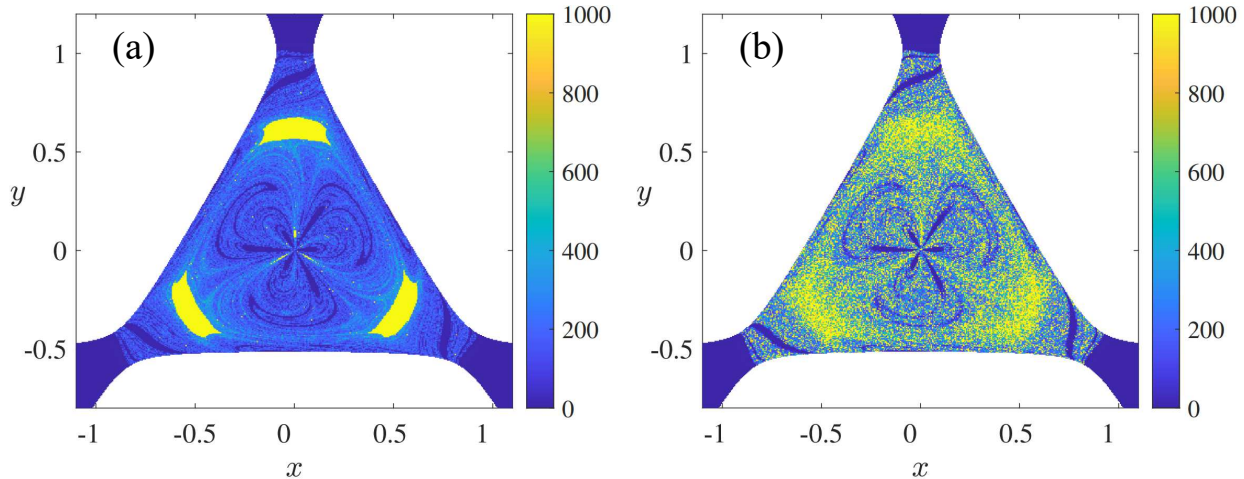}
	\caption{Average escape time distribution for the Hénon-Heiles system with energy $E=0.18$ and noise intensities (a) $\xi=10^{-10}$ and (b) $\xi_t=1.26\times10^{-6}$. To generate these figures we have computed in the grid the average escape time of every initial condition $50$ times, and calculated the average. As it can be seen in the color bar, hot colors indicate high average escape times, while cold colors refer to low escape times.}
	\label{Fig8}
\end{figure}

To illustrate the above reasoning, we have computed the escape time distributions for a very weak noise intensity $\xi=10^{-10}$ and the noise intensity that enhances the trapping for $E=0.18$ ($\xi_t=1.26\times10^{-6}$). The result can be observed in Fig.~\ref{Fig8}. For the weakly noisy case, the particles that spend long transients in the scattering region are located in the vicinity of the KAM regions of the deterministic system (see Fig.~\ref{exitbasins} for comparison). This means that the particles starting far away from KAM tori cannot enter inside them, so they will escape in similar times than in the deterministic case. The result depicted for $\xi_t$ is drastically different. The KAM regions appear blurred and an important part of the basin seems smeared by particles with high escape times. The structure of the KAM region for $E=0.18$ is contained in the high escape time region, but they are not identical. In fact, the extensive yellow region consist of different KAM regions appearing for all the values of the energy. This happens because the trajectories can not only enter the KAM tori for $E=0.18$, but all the ones that appear for different energy values.

The entrance of escaping trajectories inside the KAM tori due to the noise is a well-known result. However, the possibility to enter in KAM islands existing for lower energy values is a new phenomenon that is possible because of the drop in energy. A particular noise intensity allows the particles to enter in the structure of the KAM islands, and the fluctuations in the energy catalyze this process. This is the main mechanism that explains the noise-enhanced trapping in open Hamiltonian systems, and we will discuss it in detail in the next section. 

\begin{figure}[htp]
	\centering
	\includegraphics[width=1\textwidth,clip]{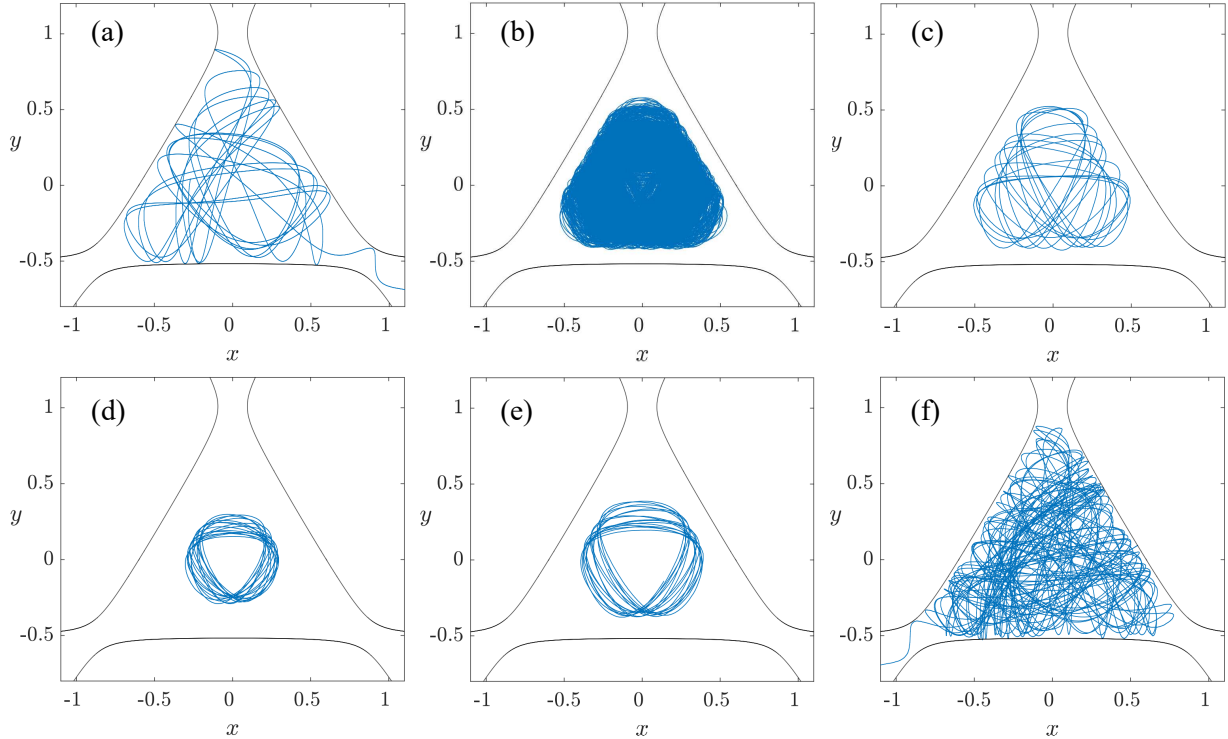}
	\caption{Evolution of two trajectories starting  in the same initial condition in the (a) deterministic and (b-f) noisy Hénon-Heiles system with energy $E=0.18$. The noise intensity is the one that enhances the trapping, $\xi_t=1.26\times10^{-6}$. Each panel shows a different time period of the evolution of the  same trajectory. In particular, the time periods are (b) $t\in[1000,10000]$, (c) $t\in[1000,1100]$, (d) $t\in[5000,5100]$, (e) $t\in[9000,9100]$, and (f) $t\in[10000,10998]$. The escape time of the deterministic case is $T=185$.}
	\label{Fig9}
\end{figure}

This relevant phenomenon can be illustrated by a simple representation in the physical space. In Fig.~\ref{Fig9} (a) we show an escaping trajectory in the deterministic system with energy $E=0.18$, escaping after a short time $T=185$. In the rest of the panels we represent a trajectory starting in the same initial condition but affected by the noise-enhanced trapping. In each panel, we show a different time period of its evolution in order to make clear that the particle does not move in a single KAM torus but in many different ones, jumping from one to another. Panel (b) shows the whole evolution of the trajectory inside different KAM tori ($t\in[1000,10000]$). Panels (c), (d) and (e) illustrate the bounded motion with different energy during the time intervals $t\in[1000,1100]$, $t\in[5000,5100]$, and $t\in[9000,9100]$ respectively. Finally, panel (f) displays the escape of the trajectory after jumping out from the KAM tori. We have carried out numerical simulations showing that this phenomenon takes place also in the hyperbolic regime of the system.  

\begin{figure}[htp]
    \centering
    \includegraphics[width=0.55\textwidth,clip]{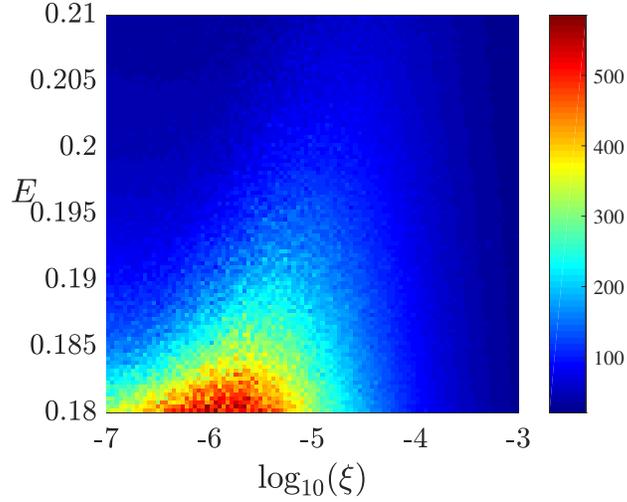}
    \caption{Color-code map showing the average escape times for several values of the energy and the noise intensity. The hot colors indicate high values of the escape times. We have used $200\times 200$ equally spaced values of the parameters. For every combination of energy and noise we have computed $50000$ trajectories in order to calculate the average escape time. The figure clearly shows that the value of the noise that enhances the trapping decreases with decreasing energy. }
    \label{pcolor_nonhy}
\end{figure}

To finalize the results in the mixed-phase-space regime, in
Fig.~\ref{pcolor_nonhy} we use again a color-coded map to show a
general portrait of the noise-enhanced trapping. The result is quite
similar to that of Fig.~\ref{pcolor_hy}. In fact,
Fig.~\ref{pcolor_hy} is a natural continuation of this figure for
higher energies.

\section{The mechanism of noise-enhanced trapping} \label{Mechanism}

\begin{figure}[htp]
    \centering
    \includegraphics[width=0.45\textwidth,clip]{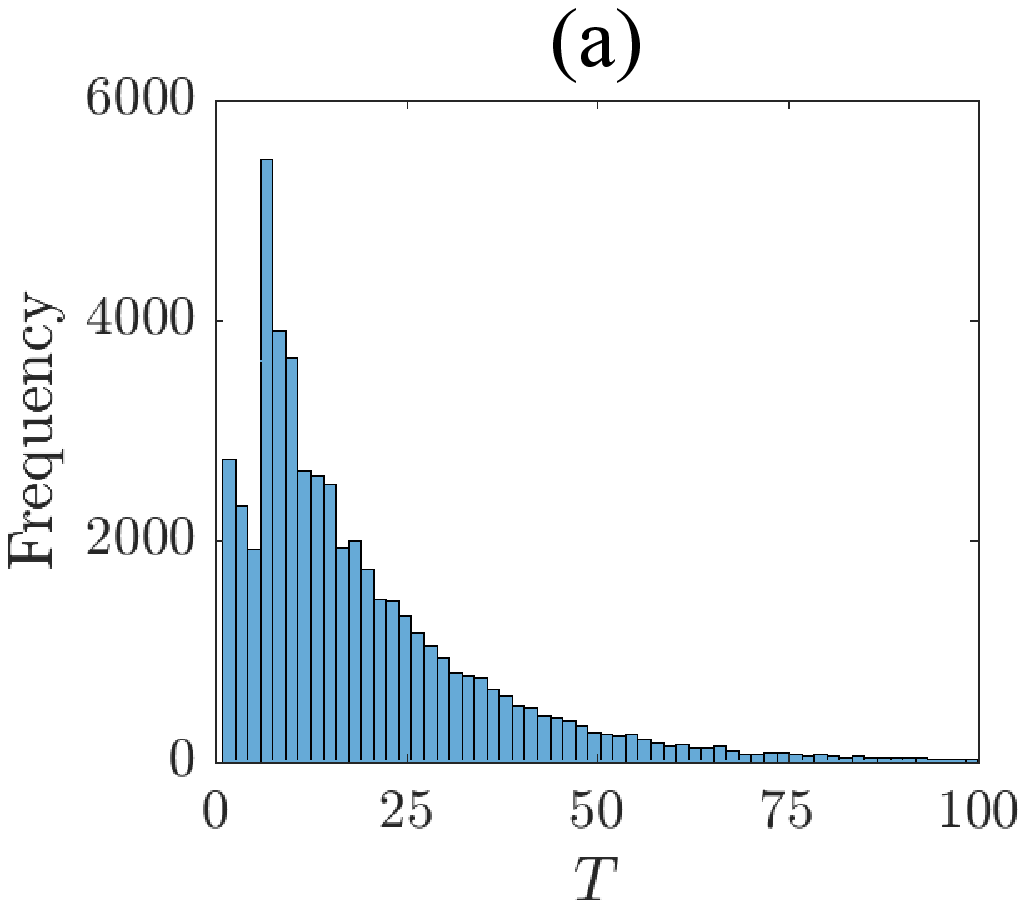}
        \includegraphics[width=0.45\textwidth,clip]{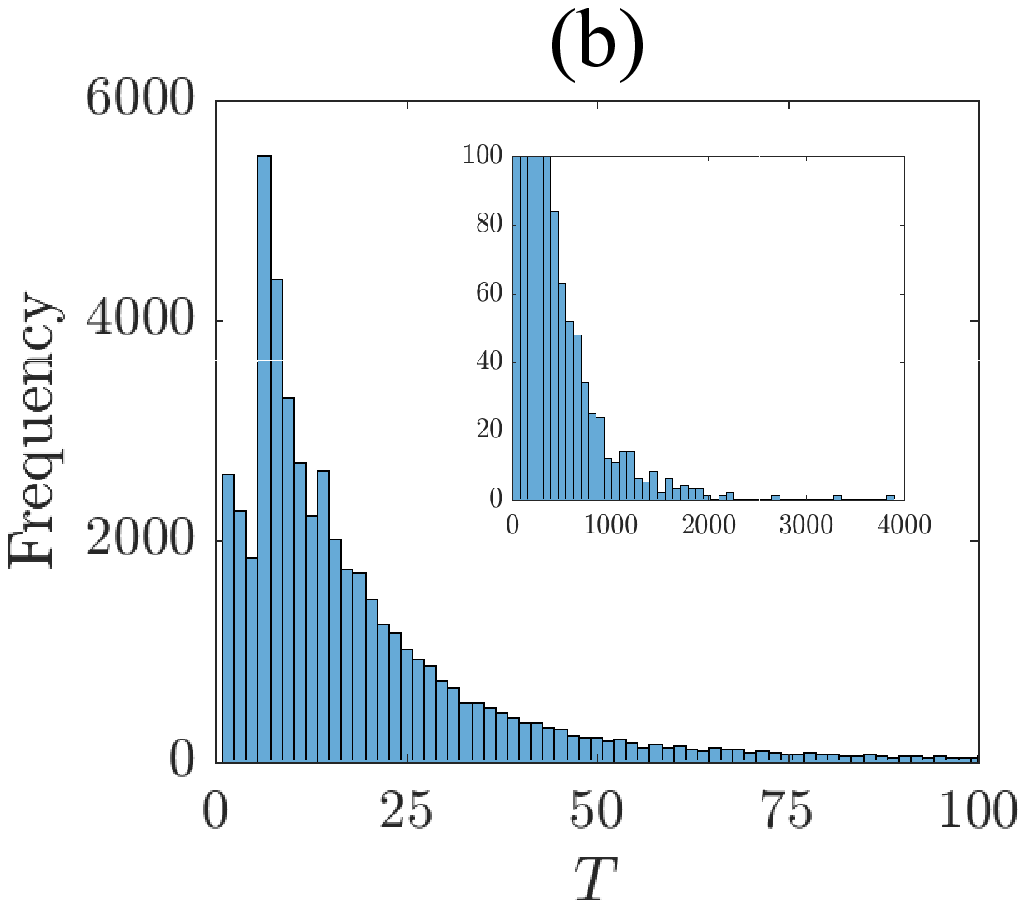}

    \caption{Histogram showing the escape time distribution for (a) $\xi=0$ and (b) $\xi_t=6.31\times10^{-5}$. The initial energy of the system is $E=0.23$. In the deterministic case the maximum escape time is $T=207$, while for $\xi_t$ a small number of trajectories survive even until $T=3887$, as shown in the inset. To generate these histograms we have launched $50000$ initial conditions in phase space and calculated the average escape time for all of them. The average escape time is $T=19.24$ for $\xi=0$ and $T=29.55$ for $\xi_t=6.31\times10^{-5}$, so the trapping is enhanced. }
    \label{hist}
\end{figure}

In the previous sections, we have shown that a noise-enhanced
trapping phenomenon appears in both the fully chaotic and
mixed-phase-space regimes. Furthermore, we have provided some
numerical evidence to show that regular trajectories of the deterministic system can enter in regions under the influence of KAM tori due to the effects of noise. These particles, that could escape in short times in the absence of noise, describe very long transients and generate a relative maximum in the evolution of the average escape time. In systems where the energy (or another relevant parameter) is not influenced by the noise, the mechanism leading to the trapping is well known, as we have discussed in the Introduction of this manuscript. However, in the context of open Hamiltonian systems, a deep analysis on the energy fluctuations is necessary to understand the trapping phenomenon. We have tested the following results with several energies in the range $E\in[0.18,0.25]$. Nevertheless, for illustrative purposes, we will show here the simulations of the noisy system with initial energy $E=0.23$. In
order to analyze what is happening with the trapping particles, we represent in Fig.~\ref{hist} a histogram showing the
escape time distribution in short times, in (a) the deterministic
case and in (b) the noisy case with $\xi_t$. Even though the results
are quite similar until $T=100$, the maximum escape time in the
deterministic case is $T=207$, while in the trapping is $T=3887$
[see inset in Fig.~\ref{hist}(b)]. Hence, we can conclude that some
unusual trajectories remain in the scattering region during very
long transient and enhance the trapping.

\begin{figure}[htp]
    \centering

    \includegraphics[width=0.45\textwidth,clip]{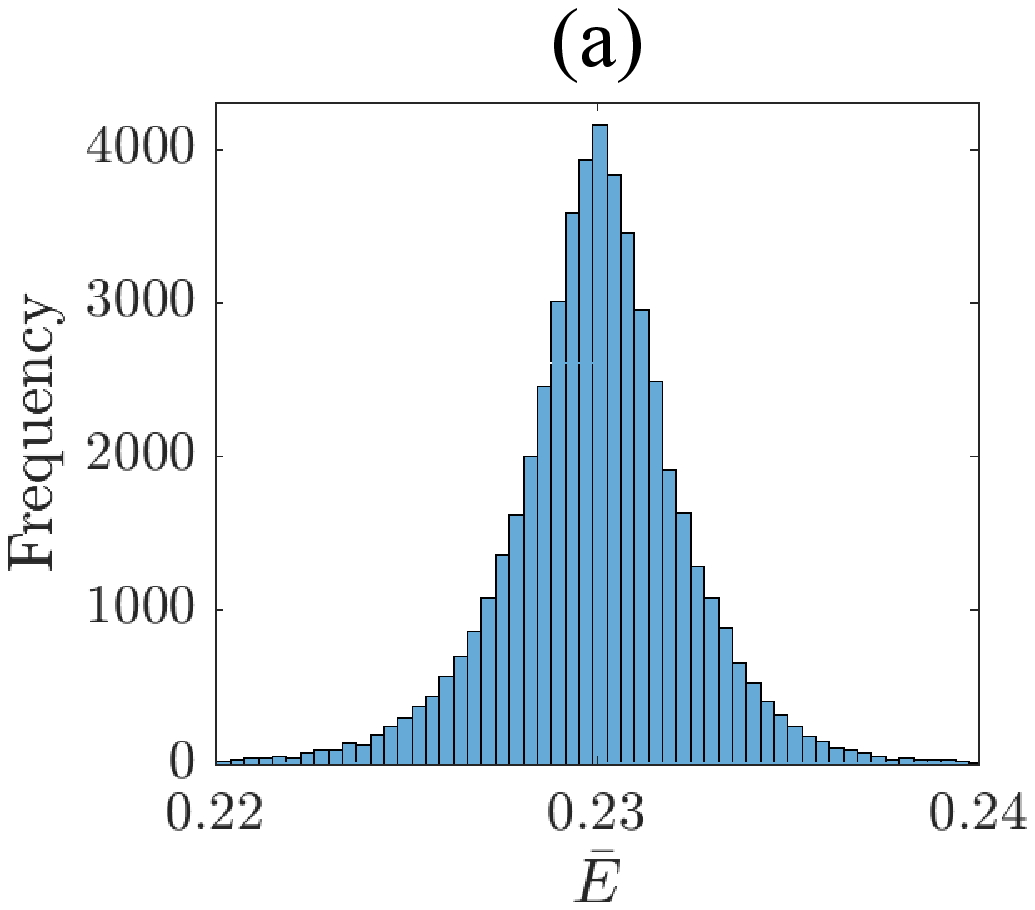}
        \includegraphics[width=0.45\textwidth,clip]{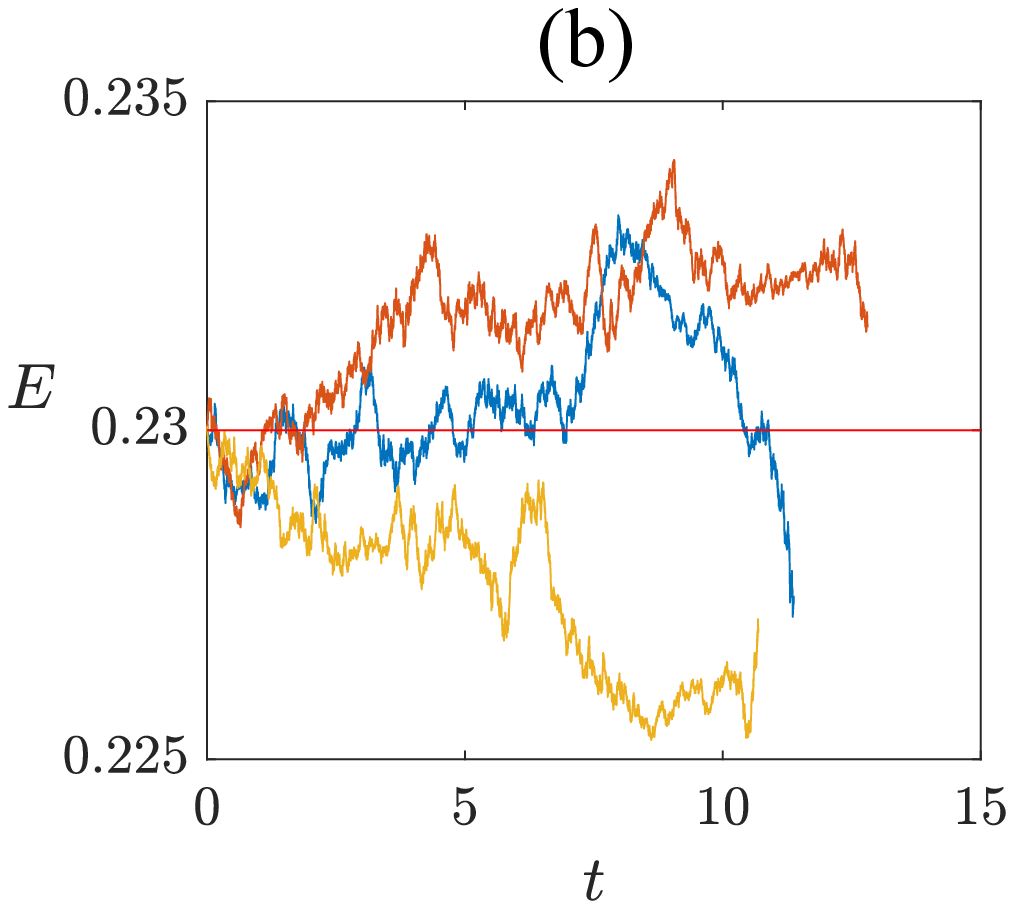}

    \caption{(a) Histogram showing the distribution of the average energies of the particles in the system with $E=0.23$ and a weak noise of intensity $\xi=10^{-6}$. (b) Three usual particles for the same values of energy and noise, escaping in short times after weak fluctuations. The red line indicates the original value of the energy.}
    \label{hist_w}
\end{figure}

We have analyzed these unusual trajectories and found the main
characteristic that differentiate them from the usual trajectories
that escape in short times: their average energy. In the
deterministic system the energy is fixed by the Hamiltonian and only
a negligible decrease (on the order $10^{-13}$ using our numerical
scheme) appears after long integration times due to the numerical
noise. However, in the presence of noise the energy is not preserved
anymore. Because of the noise, particles can move with an average
energy higher or lower than the initial one, $E_0$. Even so, in the
white noise process the mean is zero, so it makes sense to think that a
similar number of particles have an average energy $\bar{E}=E_0 +
\Delta E$ and $\bar{E}=E_0 - \Delta E$. This reasoning is valid for
low noise intensities. In Fig.~\ref{hist_w}(a) we use a histogram to
show the average energy distribution in the presence of a weak noise
intensity $\xi=10^{-6}$. We can clearly observe a Gaussian
distribution centered on $E_0=0.23$ and energies ranging from
$E=0.22$ to $E=0.24$. This means that the weak noise cannot
drastically change the average energy, leading to values that could
influence the average escape times. What happens is simply that some
trajectories have a slightly higher or lower energy, but on average
they escape with the same times as the deterministic system. As a
visual example, the evolution of the average energy for three
particles escaping at $T<15$, showing weak fluctuations in the
energy is depicted in Fig.~\ref{hist_w}(b).

\begin{figure}[htp]
    \centering
    \includegraphics[width=0.45\textwidth,clip]{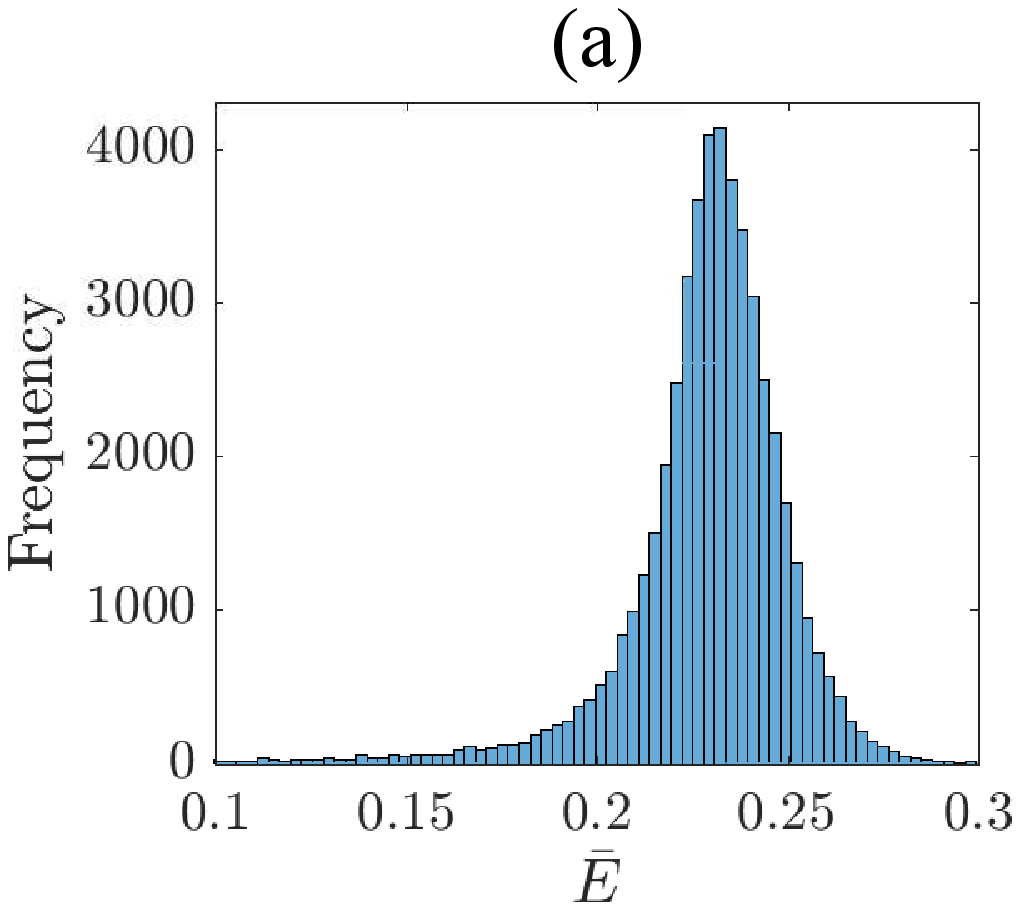}
        \includegraphics[width=0.45\textwidth,clip]{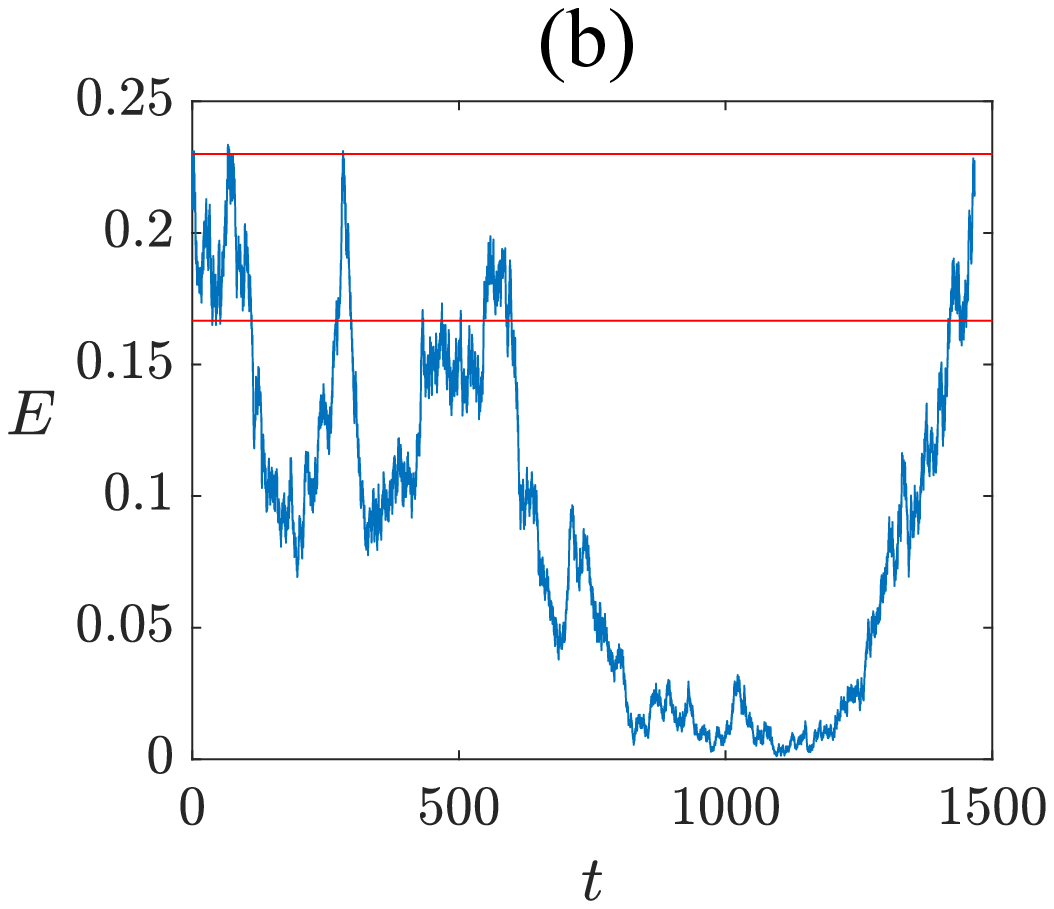}
    \caption{(a) Histogram showing the distribution of the average energies of the particles in the system with $E=0.23$ and the noise intensity that enhances the trapping, $\xi_t=6.31\times10^{-5}$. (b) An unusual trajectory remaining in the scattering region during almost $t=1500$. The upper red line indicates the original energy, while the lower refers to the threshold, $E_e$, in which the isopotential curves are opened. }
    \label{hist_t}
\end{figure}

The scene is quite different in the case of noise intensity $\xi_t$.
The fluctuations in the energy allow some particles to move in the
scattering region with a wide range of average energies. In
Fig.~\ref{hist_t}(a) the histogram shows a Gaussian distribution of
the average energies. The distribution has a maximum at $E=E_0$, as in
the case of weak noise, but the average energy ranges from $E=0.1$ to
$E=0.3$. The Gaussian is incomplete in the right side due to the
impossibility of remaining in the scattering region when having very high
energies. This distribution of average energies means that some
unusual trajectories can decrease their energy and remain in the
scattering region during long transients. These particles are the
ones that enhance the trapping. The lower the energy, the smaller
the exit set and more probable is to spend long times in the
potential. Moreover, as we have mentioned in the introduction, for
energies $E<E_e=1/6$ the exits are closed and the particles cannot
escape. In order to visualize this process, in Fig.~\ref{hist_t}(b)
the evolution of the average escape energy of one of these unusual
trajectories is represented. This particle, that should escape in a
time $T\approx 25$ in the deterministic system, decreases its
energy, moving most of the time with $E<E_e$. After a very long
transient, the fluctuations lead to an increasing in the energy and
the particle escapes in $t\approx 1500$.

In the case of low noise intensities the trajectories have not the
option of decrease noticeably the energy due the weak fluctuations.
In the trapping the fluctuations are large enough to allow some
particles to be trapped with energies $E<E_e$, or simply describe
large transients with low energies. The bigger the energy, higher
the increment $\Delta E$ necessary to reach low energy levels that
allow the particle to remain in the scattering region. This is the
reason that the value of $\xi_t$ increases with increasing values of the energy.

\begin{figure}[htp]
        \centering

    \includegraphics[width=0.55\textwidth,clip]{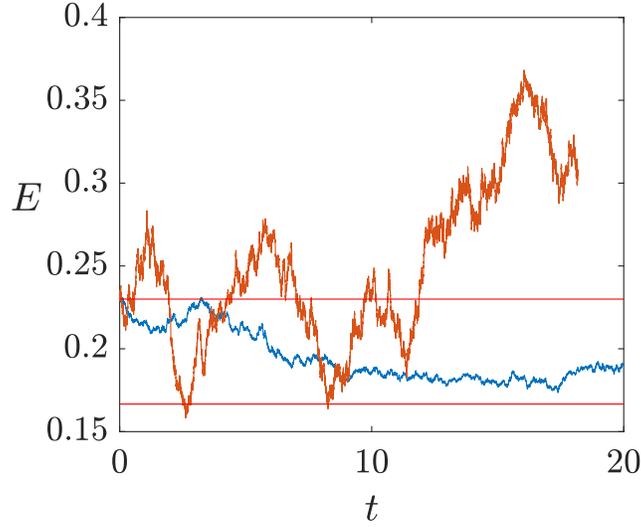}
    \caption{Comparison between the fluctuations in the energy due to a strong noise $\xi=10^{-3}$ (red curve) and due to the noise that enhance the trapping $\xi_t=6.31\times10^{-5}$ (blue curve). The upper red line indicates the original energy $E=0.23$, while the lower refers to the threshold, $E_e$, in which the isopotential curves are opened. }
    \label{compar}
\end{figure}

At this point we could ask, why in the presence of high noise the
fluctuations do not lead to an increasing of the average escape
times, but the opposite? Under the influence of high intensity
noise, the system can reach very low energy values. However, very
violent fluctuations in the energy appear and the particles
eventually escape in short times. This generates a decrease in the average escape time by increasing the
energy to very high values where the escape time of the
deterministic system is close to zero. To show this, we compare the
intensity of the fluctuations for a strong noise $\xi=10^{-3}$ and
for $\xi_t$ in Fig.~\ref{compar}. In a very short time ($t=17$) the
violent fluctuations of the strongly noisy system (red curve) reach
energies close to $0.15$ and over $0.35$, leading to an escape in
$t<20$. In the case of considering $\xi_t$ (blue curve) the
fluctuations allow the particle to avoid the escape by slowly reducing the energy. In fact, this trajectory is the same that we have
shown in Fig.~\ref{hist_t}(b), so even if we plot the evolution of the energy only until $t=20$, the trajectory remains in the scattering region until almost $t=1500$.

This mechanism related to the drop in energy explains the noise-enhanced trapping in the range of energies that we have considered ($E\in[0.18,0.25]$), which is represents both regimes in the scattering problem. The decrease in the energy opens different paths that participate in the trapping. First, the smaller the energy, the smaller the size of the exits, so the average escape time is decreased. On the other hand, for low energy values the KAM islands are bigger. Therefore, it is more probable that trajectories can jump inside one of them. Finally, in the Hénon-Heiles system and many other open Hamiltonian systems, the isopotential curves are closed under the threshold $E_e$, and hence the particles that reach such energy are trapped until a chain of positive fluctuations generates the escape. Since the main factors generating the trapping are more relevant for low energy levels, the more we increase the energy of the deterministic system, the greater the noise required to achieve the trapping. If we increase the energy more and more over $E=0.25$, we will need huge noise intensities to reach the trapping. These noise intensities could be physically meaningless.

\newpage
\section{Conclusions } \label{Conclusions}

To summarize, our research reveals that a new mechanism that enhances the trapping occurs in continuous open Hamiltonian systems. To show this, we have provided both strong numerical evidence and theoretical arguments. The numerical simulations have been based on the average escape time of the particles and the probability distributions of the energy and the escape time. In the case of the mixed-phase-space regime, we have also shown the gradual reduction of the stickiness of the KAM islands for noise intensities weaker than the noise-enhanced trapping.  

We have shown that this characteristic of chaotic scattering is
related to the influence on the escape times of some unusual
particles that decrease their energy due to the stochastic
fluctuations, leading to a very long transient. We have provided
numerical evidence by using the Hénon-Heiles Hamiltonian. However,
we expect that this result can appear, with different values of the
noise and different intensities of the trapping, in almost all
continuous open Hamiltonian systems. In particular, the so called
Barbanis potential \cite{Barrio09}, with applications to
astrophysics \cite{Contopoulos02,Navarro} and quantum mechanics
\cite{Babyuk}, and other Hamiltonians used to model galactic movements \cite{Contopoulos02,Kandrup} are perfect candidates to exhibit this
phenomenon.

We expect that this work could contribute to the general understanding of
the effects of noise in chaotic problems. We also
hope that these results can have an application in many fields of
science in which the noise models the effect of internal
irregularities or the coupling of the system with the environment.
Some examples are chaotic advection of fluids \cite{Baltanas,Daitche,Daitche2} and prey-predator competition models \cite{Prey}.

\section*{ACKNOWLEDGMENTS}
This work has been supported by the Spanish State Research Agency
(AEI) and the European Regional Development Fund (ERDF, EU) under
Projects No.~FIS2016-76883-P and No.~PID2019-105554GB-I00.

\end{document}